\documentclass[]{spie}  
\usepackage[]{graphicx}

\title{Physics of reverse annealing in high-resistivity \\ Chandra ACIS CCDs} 

\author{C. E. Grant\supit{a}, B. LaMarr\supit{a}, G. Y. Prigozhin\supit{a}, S. E. Kissel\supit{a}, S. K. Brown\supit{b} and M. W. Bautz\supit{a}
\skiplinehalf
\supit{a}Kavli Institute for Astrophysics and Space Research, Massachusetts Institute of Technology, Cambridge, Massachusetts, USA; 
\skiplinehalf
\supit{b}NASA Goddard Space Flight Center, Greenbelt, Maryland, USA
}

\authorinfo{Further author information: (Send correspondence to C.E.G.)\\C.E.G.: E-mail: cgrant@space.mit.edu}

\begin{document} 
  \maketitle 

\begin{abstract}
After launch, the Advanced CCD Imaging Spectrometer (ACIS), a focal plane instrument on the {\it Chandra X-ray Observatory}, suffered radiation damage from exposure to soft protons during passages through the Earth's radiation belts. An effect of the damage was to increase the charge transfer inefficiency (CTI) of the front illuminated CCDs. As part of the initial damage assessment, the focal plane was warmed from the operating temperature of --100$^\circ$~C to +30$^\circ$~C which unexpectedly further increased the CTI. We report results of ACIS CCD irradiation experiments in the lab aimed at better understanding this reverse annealing process. Six CCDs were irradiated cold by protons ranging in energy from 100~keV to 400~keV, and then subjected to simulated bakeouts in one of three annealing cycles. We present results of these lab experiments, compare them to our previous experiences on the ground and in flight, and derive limits on the annealing time constants.
\end{abstract}


\section{INTRODUCTION}
\label{sec:intro}

The Advanced CCD Imaging Spectrometer (ACIS), one of two focal plane science instruments on the {\it Chandra X-ray Observatory}, utilizes frame-transfer charge-coupled devices (CCDs) of two types, front- and back-illuminated (FI and BI).  Soon after launch it was discovered that the FI CCDs had suffered radiation damage from exposure to soft protons scattered off the Observatory's grazing-incidence optics during passages through the Earth's radiation belts\cite{gyp00,odell}.  The BI CCDs are less sensitive to the low energy particles which damaged the FI CCDs because the particles cannot reach the transfer channel.  In an attempt to anneal the radiation damage, the focal plane was warmed from the operating temperature of --100$^\circ$~C to +30$^\circ$~C.  Unexpectedly, the warmup had the reverse effect and caused a substantial increase in the charge transfer inefficiency.  To better understand this anomalous annealing, a laboratory experiment was developed.

Our initial experiment in low temperature irradiation and room-temperature annealing took place in September 2002 and was reported in Ref.~\citenum{bakepaper1} (Paper I).  A single CCD was irradiated with 120~keV protons while at a temperature of --100$^\circ$~C, then warmed to +30$^\circ$~C for 8~hours.  The entire irradiation/annealing cycle was designed to duplicate the experience of the flight instrument.  In addition to the expected substantial increase in CTI due to the initial irradiation, we found the warmup itself produced an additional factor of $\approx 2.5$ increase in CTI.  The room temperature bakeout also produced dramatic changes in the de-trapping times of the electron traps that cause CTI.  While the initial experiment qualitatively duplicated the ACIS on-orbit experience, the relative magnitude of the reverse annealing was larger in the ground data by a factor of four.

We speculated that the room-temperature annealing changed the trap population, converting traps with relatively benign re-emission time constants to more harmful ones.  In particular, the ACIS CCDs contain carbon impurities.  Irradiation of the CCDs creates vacancies and silicon interstitials which are highly mobile at ACIS operating temperatures.  The silicon interstitials can exchange places with the carbon forming a carbon interstitial ($C_i$) which is not mobile at ACIS operating temperatures and does not affect CTI.  At room temperatures, however, the carbon interstitials are free to form defects that can negatively affect CTI, such as $C_i$-$P_s$.
 
In this paper we describe a follow-up experiment, undertaken in May 2005, with multiple CCDs, irradiation/annealing profiles, and proton energies to better explore the physics behind the anomalous annealing observed in Paper I.  Our goals were to understand why our ground test results differed from the on-orbit bakeout experience, to test our model of the ``reverse annealing'' phenomenon and to gain greater confidence in predicting CCD changes resulting from any future ACIS bakeout.  By irradiating multiple CCDs we can determine the importance of chip-to-chip variations.  The energy dependence of annealing can be tested using multiple proton energies and different annealing profiles can be used to study the timescales of the phenomenon.  Section~\ref{sec:exper} describes the experimental setup and charge transfer inefficiency data analysis.  Section~\ref{sec:results} examines the properties of the anomalous annealing, while section~\ref{sec:disc} compares these results to our previous experiment.

\section{EXPERIMENTAL DETAILS}
\label{sec:exper}

The ACIS CCDs are buried channel devices made of high-resistivity float zone silicon with a depletion depth of 50--75~$\mu$m.  Each CCD is a framestore-transfer device with 1024 by 1026 pixels in the imaging and framestore arrays.  The pixels in the imaging array are 24~$\mu$m square.  The image-to-framestore transfer rate is 40~$\mu$s per row and the pixels are read out in four output nodes at $10^5$~pixels/s.  In total six CCDs were irradiated and annealed; five from the backup ACIS focal plane and one from an ACIS flight production lot that was used in our initial experiment (Paper I).  Only two of four quadrants of this CCD were previously irradiated.

The irradiating protons at energies of 100, 120, 180, and 400~keV were provided by the 2~MeV van de Graaff accelerator at the Goddard Space Flight Center radiation laboratory.  A surface barrier detector inserted between the accelerator and our test setup was used for dosimetry.  The radiation dose was chosen to cause charge transfer inefficiency of $\sim 10^{-4}$, to approximately match the irradiation of the flight CCDs.  The dose rate was typically $\sim 5 \times 10^{4}$ p/cm$^{2}$/s for a total dose of $2 \times 10^7$ to $2 \times 10^8$ protons/cm$^{2}$ at the monitor, depending on energy.  The CCDs were irradiated cold, at $-100^o$C, and powered off to duplicate the flight environment during radiation belt transits.

\begin{figure}
\vspace{4in}
\includegraphics{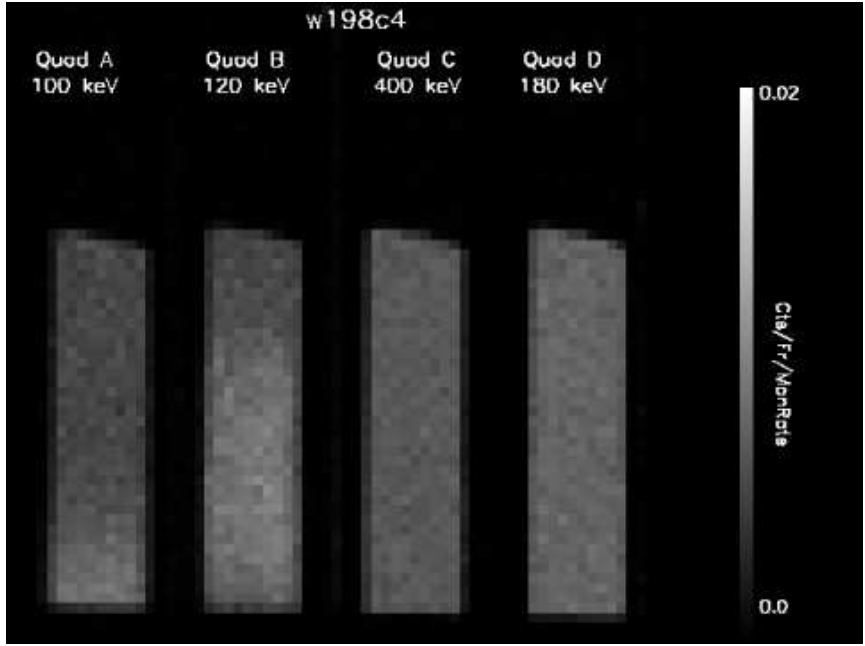}
\caption{An example of the low-flux images that were used to align the proton beam.  Four images of different energy proton beams are summed here to show the irradiated region in each quadrant of the CCD w198c4.  The CCD framestore and readout are at the bottom of the image.}
\label{fig:radim}
\end{figure}

The camera held two CCDs, side by side, utilizing the same electronics as the ACIS flight instrument.  A slit-shaped baffle in the beam confined the protons to an approximately 3.5~mm by 12~mm region, aligned to fit within a single CCD readout quadrant.  Like the flight devices, the framestore and serial transfer arrays were shielded during irradiation and were therefore undamaged.  A flexible bellows was used to attach the camera chamber to the beam line which, along with a movable camera table, allowed the CCD chamber to pivot placing any of the eight CCD quadrants in the proton beam without exposing the other quadrants.  The CCD was aligned by directly imaging a low flux proton beam with the CCD itself.  An example of the low flux beam snapshots is shown in Figure~\ref{fig:radim} and shows both the uniformity and the well-defined edges of the irradiated regions.

A movable $^{55}$Fe source could be inserted and was used to monitor charge transfer inefficiency (CTI) before and after each irradiation and annealing.  The CCDs were cooled to --100$^\circ$~C, the nominal focal plane temperature during the flight annealing experience, during all CTI measurements.  To measure CTI, the CCDs were uniformly illuminated by the $^{55}$Fe source which produces a strong spectral line of Mn-K$\alpha$ at 5.9 keV.  The low-flux proton beam measurements were used to determine the boundaries of the irradiated regions on the CCD.  For CTI measurement, the event lists were filtered to include only the events in these regions.  The exposure times were set to accumulate $\sim$ 1.5 million counts, so each data set has similar statistics.  Parallel CTI was measured by fitting a linear function to the center pixel pulseheight as a function of row.  Serial CTI is unaffected by the low-energy protons used here and was not measured.

Three different irradiation/annealing profiles were used.  In all cases, the CCDs were powered off during both irradiation and annealing.  The first, intended to duplicate both the flight experience and our earlier experiment, was a cold irradiation followed by an 8-hour annealing at +30$^\circ$~C.  The second was a multi-temperature isochronal annealing cycle to determine the sensitivity of the annealing rate to temperature.  After irradiation, the CCDs were annealed for one hour each at 0$^\circ$~C, +10$^\circ$~C, +20$^\circ$~C, and +30$^\circ$~C.  The temperature was returned to --100$^\circ$~C between each anneal to monitor CTI.  A third long duration annealing was done at the conclusion of the isochronal annealing cycle.  The CCDs were annealed at +30$^\circ$~C for one hour, one hour again, six hours and a final anneal of over 100 hours.  Again, the temperature was returned to --100$^\circ$~C between each anneal to monitor CTI.

\section{RESULTS}
\label{sec:results}

We then use our measured CTI to examine the properties of the reverse annealing.  Before we can understand our results, however, we need to confront two sources of systematic error and either correct the CTI data or increase the error associated with each measurement to compensate.  These error sources are due to variable temperatures and to post-irradiation CTI relaxation.

\subsection{Temperature Variation Corrections}
\label{sec:temp}

\begin{figure}
\vspace{3.7in}
\includegraphics{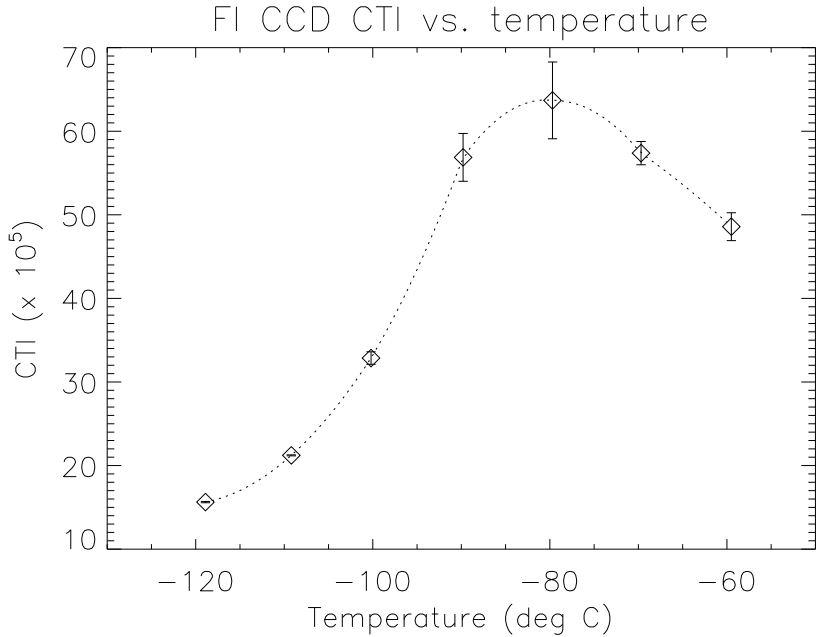}
\caption{Parallel CTI versus temperature for an ACIS FI CCD on orbit.  The dotted line is an interpolation intended to highlight the curvature.  The CTI of the radiation damaged ACIS CCDs has a strong dependence on temperature at --100$^\circ$~C.  Taken from Ref.~\citenum{ctitemp}}
\label{fig:temp}
\end{figure}

Measured CTI is a function of temperature.  Detrapping time constants decrease as the temperature increases so that different populations of traps can become more or less important.  If the detrapping time constant drops below the pixel-to-pixel transfer time (40 $\mu$s for the ACIS imaging array) or becomes much longer than the typical time between charge packets during transfer, charge is no longer lost to that particular trap.  The distribution of trap time constants at a particular temperature determines the CTI, so temperature can positively or negatively correlate with CTI.  The CTI of the radiation damaged ACIS CCDs has a strong dependence on temperature of nearly 10\% per degree at --100$^\circ$~C\cite{ctitemp}.  Figure~\ref{fig:temp} shows a measurement of the CTI-temperature dependence for the ACIS CCDs in orbit.  Effort was made both to keep the temperature stable during individual measurements and to keep the temperature repeatable over the weeks of the experiment, however this was not entirely successful.  

The CCD temperature was regulated by a liquid nitrogen cooled cold block, RTDs and a closed-loop heater.  The cold block temperature was controlled by an open-loop, timed LN$_2$ valve.  The duty cycle of the LN$_2$ valve was not identical for all measurements which led to small CCD temperature differences.  Where possible, we have applied a post-facto correction to the measured CTI to account for the temperature variation.

We assume that further annealing after the initial 8-hour +30$^\circ$C anneal does not change CTI (this will be further explored in Section~\ref{sec:long}) and that the fractional change in CTI with temperature is identical for all CCDs.  Based on our experience with the flight instrument, the fractional change in CTI with temperature is roughly the same for FI CCDs within a few tenths of a percent.  Since data was always taken with two CCDs simultaneously, we can, when possible, use the already-annealed CCD quadrants to track any temperature changes.  It is possible to apply such a temperature-correction in 15 of 19 annealing measurements.  The size of the correction varies, from 1\% to as much as 25\%.  For the remaining measurements, where correction is not possible, we adopt a systematic error of 10\% in CTI.

\subsection{Post-Irradiation CTI Relaxation}

\begin{figure}
\vspace{3.6in}
\includegraphics{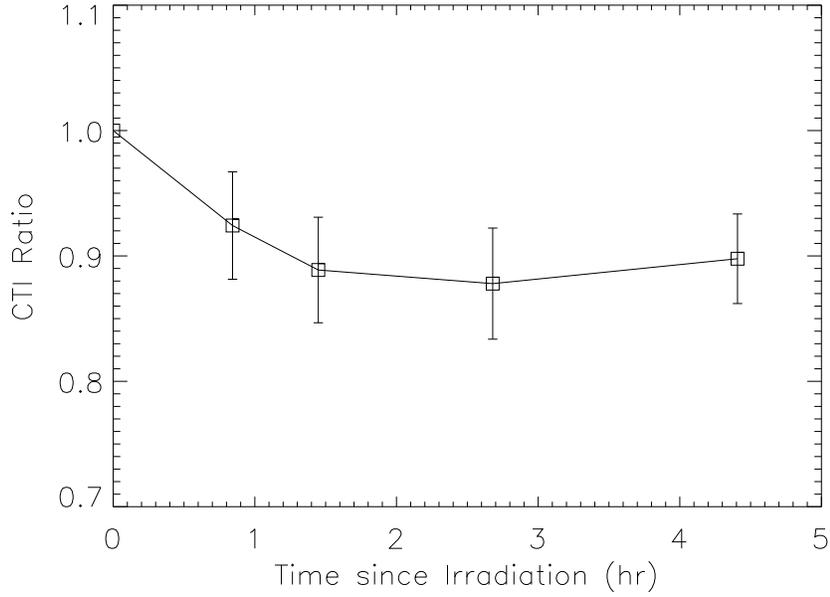}
\caption{Ratio of pre-annealing CTI measurements to the CTI measured immediately after irradiation as a function of time.  The data from multiple CCD quadrants have been binned by time to produce this average CTI relaxation profile.  Four hours after irradiation the mean fractional change is $-10\% \pm 4\%$.}
\label{fig:relax}
\end{figure}

An additional unexpected source of systematic error is that repeated CTI measurements immediately after irradiation, but before annealing, show clear ``relaxation'' or reduction of CTI during the first hour after irradiation.  For most CCD quadrants, annealing took place many hours after irradiation, and multiple pre-annealing CTI measurements were made, so we can determine the magnitude of the effect and then use this information to correct for it.  Figure~\ref{fig:relax} shows the average CTI change as a function of time after the first post-irradiation CTI measurement. The mean fractional change four hours after irradiation is $-10\% \pm 4\%$.  We apply this as a correction to the data immediately following each irradiation.

\subsection{Anomalous Annealing Results}

\begin{figure}
\vspace{3.6in}
\includegraphics{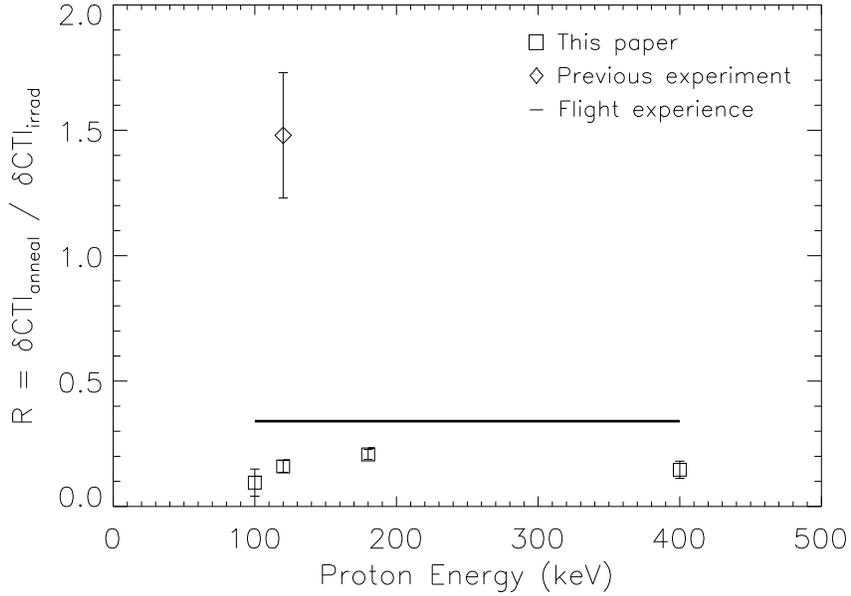}
\caption{The fractional increase in CTI due to annealing, $R$, as a function of proton energy.  The diamond and squares indicate data from our original annealing experiment (Paper I) and from this paper.  The solid horizontal line indicates the fractional increase from annealing as measured by the flight instrument.  The error bars indicate 90\% confidence levels.}
\label{fig:R}
\end{figure}

In this experiment, as in our previous experiment and the flight experience, room-temperature annealing increases the CTI of the CCDs.  If we assume that the CTI increase due to annealing should be proportional to the CTI increase from the pre-anneal irradiation independent of dose, then we can define the ratio, $R$, such that $$R = \frac{\delta CTI_{annealing}}{\delta CTI_{irradiation}}.$$  The measured value of $R$ for each CCD quadrant individually and averaged over all proton energies for each CCD is shown in Table~\ref{tab:R} along with the results from our previous experiment in 2002 and the flight experience.  The mean value of $R$ as a function of proton energy is shown in Figure~\ref{fig:R}.  There is some indication of a weak dependence of $R$ on energy, increasing from 100 to 180~keV, then decreasing to 400~keV.

\begin{table}
\caption{Fractional Increase in CTI after 8-hour +30$^\circ$~C Annealing}
\label{tab:R}
\begin{center}
\begin{tabular}{lrrrrrrrrr}

CCD ID &\multicolumn{4}{c}{Proton Energy (keV)} &\multicolumn{4}{c}{$R$} &\multicolumn{1}{c}{$<R>$} \\ 

 &\multicolumn{1}{c}{A} &\multicolumn{1}{c}{B} &\multicolumn{1}{c}{C} &\multicolumn{1}{c}{D} &\multicolumn{1}{c}{A} &\multicolumn{1}{c}{B} &\multicolumn{1}{c}{C} &\multicolumn{1}{c}{D} \\ \hline \\

w215c1 &$\cdots$ &120 &400 &180 &\multicolumn{1}{c}{$\cdots$} &0.41 (0.23) &0.21 (0.07) &0.23 (0.03) &0.22 (0.03) \\
w216c3 &$\cdots$ &120 &400 &180 &\multicolumn{1}{c}{$\cdots$} &0.11 (0.05) &0.17 (0.12) &0.16 (0.04) &0.14 (0.03) \\
w192c1 &100 &120 &400 &180 &0.24 (0.10) &0.16 (0.14) &0.10 (0.16) &0.23 (0.23) &0.24 (0.10) \\
w198c4 &100 &120 &400 &180 &-0.08 (0.10) &0.18 (0.04) &0.13 (0.07) &0.23 (0.07) &0.15 (0.03) \\
w210c3 &180 &120 &400 &$\cdots$ &0.23 (0.07) &0.19 (0.05) &0.11 (0.05) &\multicolumn{1}{c}{$\cdots$} &0.17 (0.03) \\
w183c3 (2005) &$\cdots$ &$\cdots$ &100 &120 &\multicolumn{1}{c}{$\cdots$} &\multicolumn{1}{c}{$\cdots$} &0.12 (0.09) &0.16 (0.06) &0.15 (0.05)\\
\\
w183c3 (2002) &\multicolumn{2}{c}{120} & & & & & & &1.48 (0.25) \\
\\
Flight ACIS-S2 & & & & & & & & &0.34 (0.02) \\

\end{tabular}
\end{center}
\small
{Note: The error bars indicate 90\% confidence levels.}
\normalsize
\end{table}

While in all cases the room temperature annealing produces an increase in CTI, the magnitude of the increase is quite different between our original experiment and the current one and somewhat different from the flight experience.

\subsection{Long Duration Annealing}
\label{sec:long}

\begin{figure}
\vspace{3.6in}
\includegraphics{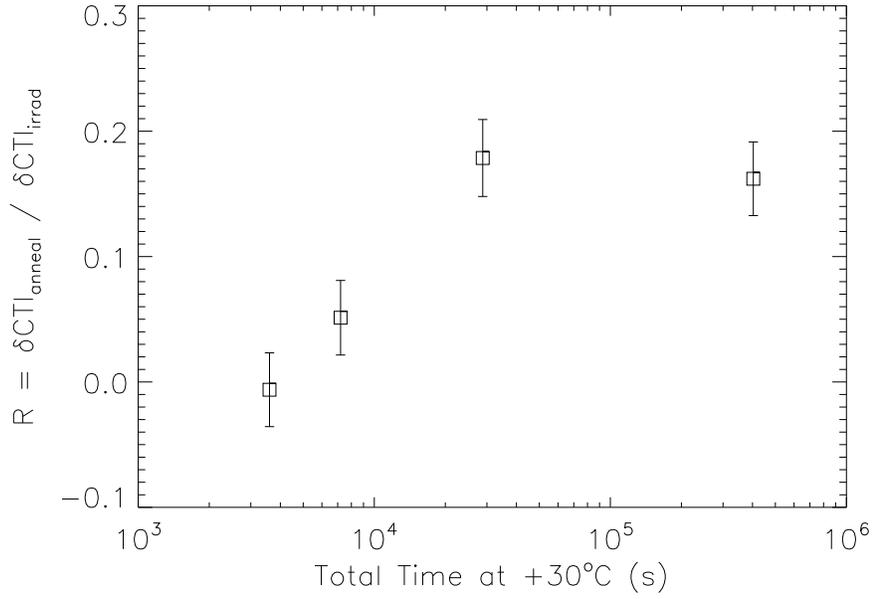}
\caption{The fractional increase in CTI due to annealing, $R$, as a function of annealing time.  The figure shows the average of four CCD quadrants which were treated to progressively longer anneals at $+30^o$C.  The error bars indicate 90\% confidence levels.  The process which increases CTI appears to be completed after eight hours.}
\label{fig:long}
\end{figure}

A long duration anneal was done at the conclusion of the isochronal annealing (which will be discussed in Section~\ref{sec:iso}).  The CCDs were annealed at a temperature of +30$^\circ$~C for intervals of one hour, one hour again, six hours and finally over a hundred hours.  Between each annealing interval, the temperature was lowered to --100$^\circ$~C to measure the CTI.  The fractional increase in CTI as a function of total annealing time is shown in Figure~\ref{fig:long}.  The first 1-hour anneal caused minimal CTI increase.  After two hours of annealing time, the CTI had increased significantly.  The maximum CTI was reached after 8 hours of total annealing time, with no additional increase even after over 100 hours of annealing time.  The time constant for annealing at +30$^\circ$~C must be more than one hour, but less than eight hours.  This is consistent with the longest of the relaxation timescales for various states of the $C_i$ - $P_s$ complex, believed to be the cause of the annealing CTI increase.  That time constant is of order $2 \times 10^4$ seconds, just shorter than the eight hour annealing time.  Our earlier assumption, in Section~\ref{sec:temp} on temperature-correction of CTI, that further annealing after the initial 8-hour +30$^\circ$C anneal does not change CTI, appears to be well founded.

\subsection{Multi-Temperature Isochronal Annealing}
\label{sec:iso}

\begin{figure}
\vspace{3.6in}
\includegraphics{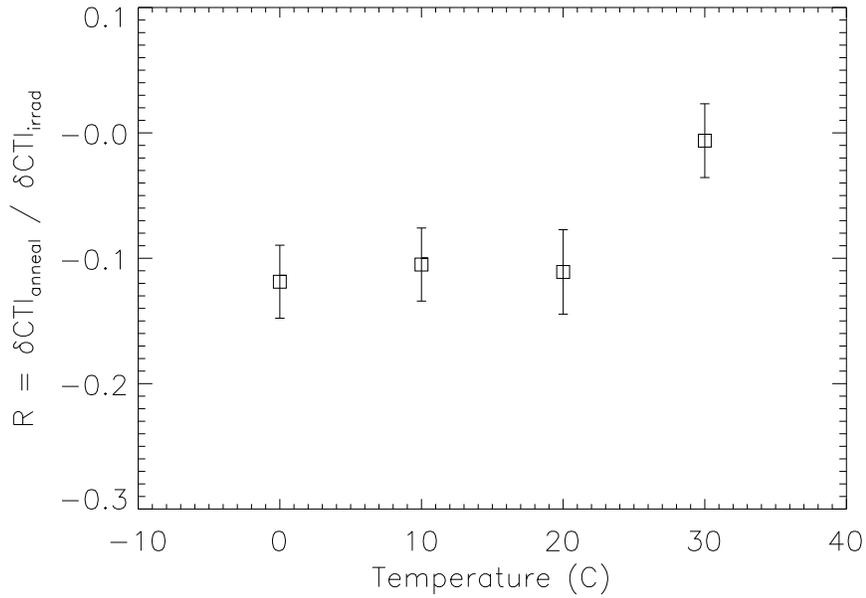}
\caption{The fractional increase in CTI due to annealing, $R$, as a function of annealing temperature.  The figure shows the average of four CCD quadrants which were treated to one hour annealing intervals at successively increasing temperatures.  The error bars indicate 90\% confidence levels.}
\label{fig:iso}
\end{figure}

A multi-temperature isochronal annealing cycle was done to test the sensitivity of the annealing rate to temperature.  After irradiation, the CCD was annealed for one hour intervals each at 0$^\circ$~C, +10$^\circ$~C, +20$^\circ$~C, and +30$^\circ$~C.  The temperature was returned to --100$^\circ$~C between each annealing interval to monitor CTI.  The long duration annealing was done at the conclusion of the isochronal annealing cycle.  From the long duration annealing test, we found that little CTI increase occurred in the initial hour of the anneal, so we might expect to see minimal change in the CTI from these short multi-temperature anneals.  Figure~\ref{fig:iso} shows the fractional increase in CTI as a function of annealing temperature.  CTI remains essentially constant through the lower temperature anneals and only increases after the annealing temperature reaches +30$^\circ$~C.  

The CTI measurement after the initial 0$^\circ$~C annealing shows a {\it decrease} in CTI and therefore a negative value of $R$.  We cannot identify a specific systematic error that could cause this CTI drop.  The final data point, at +30$^\circ$~C, in Figure~\ref{fig:iso} is also the first data point in the long duration annealing test in Figure~\ref{fig:long}.  The final value of $R$ after the long duration +30$^\circ$~C annealing is consistent with that found in the other annealing tests, so we do not think there is a systematic error shifting all of the $R$ measurements to lower values.  It is possible that there is an additional change in the trap population that initially decreases CTI at lower temperatures before increasing CTI at +30$^\circ$~C, but it is difficult to determine with our current CTI results.

\section{DISCUSSION}
\label{sec:disc}

In our initial experiment reported in Paper I, we found a substantial difference in the magnitude of the annealing-induced CTI between the ground and flight devices.  Several explanations were proposed, such as differences in impurity concentrations between CCDs and in the energy spectrum of the irradiating particles.  The multiple CCDs and proton energies utilized in the experiment described here can address both these factors.

Six CCDs were put through essentially identical irradiation and annealing cycles.  The measured fractional increase in CTI from annealing, $R$, averaged over all CCDs is 0.17 $\pm$ 0.01 (90\% confidence) with a standard deviation of 0.09.  The CCD-to-CCD variation in $R$ is larger than the expected statistical variation, so may represent real differences in the CCD composition.  As discussed in Paper I, we expect the value of $R$ to vary linearly with the concentration of carbon impurities.  Measurements of ACIS sibling devices show an RMS variation of 15\% with a maximum variation of a factor of 1.6 in the carbon concentration.  This is smaller than the measured variation in $R$, but is roughly consistent.

Variations in CCD properties cannot, however, explain the large difference between the original laboratory results and the current results.  Different regions of one CCD, w183c3, were irradiated and annealed in both experiments with highly discrepant results.  There was no significant difference between the results for w183c3 and the other five CCDs in the current experiment.  Clearly the inconsistency between the two ground experiments cannot be blamed on variations between CCDs, so must be due to other factors.

Differences in the irradiating proton energy were also suggested as reasons for the discrepancy between the original ground experiment and the flight results.  Our initial experiment used a single proton energy, 120~keV, while the current one utilized proton energies of 100, 120, 180 and 400~keV.  The flight devices were irradiated by a broad spectrum of low energy protons\cite{gyp00}.  While there does appear to be some variation in $R$ with proton energy, the magnitude is nowhere near large enough to explain the large difference between the original laboratory experiment and either the flight results or the current experiment.

If the discrepancy between the original experiment and the current one is not due to device differences or to proton energy, we must conclude that something about the experimental protocol was different between the two.  An obvious possibility is CCD temperature.  As was discussed in Section~\ref{sec:temp}, the CTI of the radiation damaged ACIS CCDs has a strong dependence on temperature of nearly 10\% per degree at --100$^\circ$~C.  The CTI measurements of w183c3 indicate that the CCD temperature in the original experiment was about 5$^\circ$~C warmer than in the current one.  This was most likely due to the different placement of the temperature sensors on the CCD camera.  If the temperature dependence of CTI before and after annealing is different, then the fractional change in CTI due to annealing would also be temperature dependent.  The distribution of trap time constants at a particular temperature determines CTI and the CTI temperature relation.  Since our explanation for the annealing CTI increase is a change in the trap population, it seems reasonable to conclude that the discrepancy between the two laboratory experiments can be assigned to temperature differences.

\section{CONCLUSION}

We report the results of laboratory experiments of cold irradiation and room-temperature annealing on ACIS CCDs.  Six CCDs were irradiated cold, at --100$^\circ$~C, by low-energy protons with energies from 100 to 400~keV.  The CCDs were then annealed at +30$^\circ$~C in one of three annealing cycles.  We used the fractional CTI increase due to annealing, $R$, to study the dependence on proton energy and intrinsic properties.  The measured value of $R$, averaged over all CCD quadrants was 0.17 with a standard deviation of 0.09.  This variation is larger than, but roughly consistent with the variance in concentration of carbon impurities in the ACIS CCDs.  We found a mild dependence on proton energy, increasing from 100 to 180~keV, then decreasing to 400~keV.  The long duration annealing tests showed that the processes which increase CTI appear to be complete after eight hours.  The results were generally consistent with ACIS flight experience, but highly discrepant with our original laboratory experiment.  We believe this is due to differences in CCD temperature.

To better understand the characteristics of the traps that cause the annealing-induced CTI increase, we plan to continue study of these data.  The results of the multi-temperature isochronal annealing, with an apparent CTI {\it decrease} at the lower temperatures (0$^\circ$~C to +20$^\circ$~C) before the final increase at +30$^\circ$~C, are puzzling and we do not currently have an explanation for this behavior.  It is likely that study of the charge in the pixels trailing the X-ray events may prove illuminating, as it was in Paper I, indicating changes in the re-emission timescales of the trap population at each annealing stage.  The trailing charge will also help further explore the differences and similarities between the two ground experiments and the flight case.

\acknowledgments
The authors would like to thank C. Smith of NASA's Goddard Space Flight Center for his assistance at the radiation laboratory.  This work was supported by NASA contracts NAS 8-37716 and NAS 8-38252. 

\bibliography{paper}   
\bibliographystyle{spiebib}   

\end{document}